\documentclass[prl,twocolumn,showpacs,floats,showkeys,superscriptaddress]{revtex4}

\usepackage{graphicx}
\bibpunct{}{}{,}{s}{}{}
\begin{document}

\title{Fast nuclear spin hyperpolarization of phosphorus in silicon}

\author{D. R. McCamey}\email{dane.mccamey@physics.utah.edu}
\affiliation{Department of Physics, University of Utah, 115 South
1400 East Rm 201, Salt Lake City, Utah 84112}

\author{J. van Tol}
\affiliation{Center for Interdisciplinary Magnetic Resonance,
National High Magnetic Field Laboratory at Florida State University,
Tallahassee, Florida 32310, USA}
\author{G. W. Morley}
\affiliation{London Centre for Nanotechnology, Gower Street, London
WC1E 6BT, United Kingdom }
\author{C. Boehme}\email{boehme@physics.utah.edu}
\affiliation{Department of Physics, University of Utah, 115 South
1400 East Rm 201, Salt Lake City, Utah 84112}

\date{\today}


\begin{abstract}
We experimentally demonstrate a method for obtaining nuclear spin
hyperpolarization, that is, polarization significantly in excess of
that expected for a thermal equilibrium. By exploiting a modified
Overhauser process, we obtain more than 68\% nuclear
anti--polarization of phosphorus donors in silicon. This
polarization is reached with a time constant of $\sim 150$ seconds,
at a temperature of 1.37 K and a magnetic field of 8.5 T. The
ability to obtain such large polarizations is discussed with regards
to its significance for quantum information processing and magnetic
resonance imaging.
\end{abstract}



\maketitle

Phosphorus doped crystalline silicon (Si:P) is a model system for
investigating spin effects in the solid state and at the same time
is a point defect with great technological importance. Si:P has been
used since the beginning of the semiconductor industry in the early
1950's for applications ranging from the ubiquitous (thin film
transistors) to the conceptual (single electron transistors). The
ability to hyperpolarize the spins in this material is important for
a number of its applications. Utilizing the nuclear spin of
phosphorus donors as quantum bits \cite{kane,Morton2008} relies on
the ability to obtain a well characterized initial state
\cite{DiVincenzo2000}, which can be obtained by hyperpolarization.
Spin polarized silicon microparticles may also have applications for
magnetic resonance imaging techniques\cite{Dementyev2008}, similar
to other hyperpolarized systems, such as xenon \cite{Schroder2006}.
Whilst it is reasonably simple to obtain large electron spin
polarization, for example by using moderate magnetic fields at
liquid $^4$He temperatures, doing the same with nuclear spins is
difficult due to their much smaller Zeeman splitting. There are a
number of schemes used to obtain nuclear spin polarization in excess
of the thermal polarization. Dynamic nuclear polarization using
off--resonance radiation has been studied extensively
\cite{Abragam1978,Dementyev2008}. Complex pulses or adiabatic
passage effects may be used to manipulate spin states, leading to
large polarizations\cite{Feher1959a, Morley2007}. Electrical
injection of hot carriers has been used to obtain positive
polarizations \cite{Clark1963}, however this requires electrical
contact to the sample. Optical excitation with linearly polarized
sub-bandgap light has given small ($\sim 2.5$\%) polarization of
$^{29}$Si nuclei in silicon with a natural isotopic abundance
\cite{Verhulst2005}.

In this letter, we demonstrate anti--polarization of phosphorus
donor nuclei in silicon of up to $P = -68\%$. The scheme used is
simple, fast and does not involve resonant manipulation of either
the nuclear or electronic spin. Instead, the relative populations
are modified using photoexcited carriers, generated using white
light, at low temperatures (about $^4$He temperature) and in
magnetic fields ($\sim 8.5$ T) significantly smaller than those
required to obtain an equivalent thermal nuclear spin polarization.

\begin{figure}[b]
\centering\includegraphics[width=7cm]{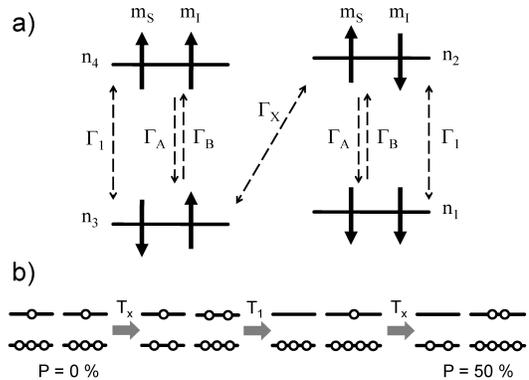}
\caption{\label{fig:RateDiagram}a) Sketch of the energy levels of
the four spin eigenstates of a phosphorus donor atom in silicon in
presence of very high magnetic fields. The dashed arrows indicate
allowed transitions with their respective rate coefficients.
$\Gamma_1$ is for longitudinal relaxation processes, $\Gamma_A$ and
$\Gamma_B$ for scattering with conduction electrons and $\Gamma_X$
for the Overhauser flip--flop process. b) Sketch of the change from
a thermally polarized spin ensemble to a hyperpolarized spin
ensemble for $T_{\mathrm{res}}>>T_{\mathrm{spin}}$. Note that the
$T_X$ and $T_1$ processes act continuously (not sequentially as
illustrated here).}
\end{figure}

Phosphorus in silicon can be described by the spin ($S=1/2$) of its
donor electron that is coupled to the spin ($I=1/2$) of the $^{31}$P
nucleus. This model provides a system with four energy levels, as
shown in Fig. \ref{fig:RateDiagram} for the presence of strong
magnetic fields when the nuclear Zeeman splitting exceeds the
nuclear to donor electron hyperfine interaction. At $B_0 \approx
8.5$ T, the donor electron Zeeman splitting is $\Delta E_\mathrm{e}
\approx 240$ GHz whereas the nuclear Zeeman energy is $\Delta
E_\mathrm{n} \approx 147$ MHz and the hyperfine interaction $A =
117$ MHz.
Figure \ref{fig:RateDiagram}(a) shows the relevant spin relaxation
processes that occur in the $^{31}$P donor atom. $\Gamma_1$ is the
rate coefficient associated with longitudinal relaxation of the
electron magnetization towards thermal equilibrium with the crystal
lattice at temperature $T_\mathrm{spin}$. $\Gamma_X$ is the rate
coefficient associated with the Overhauser spin relaxation process
(a flip-flop) between the electron and nuclear spins
\cite{Overhauser1953}. The dependence of the Overhauser rate on
temperature and magnetic field has been described by Pines et al.
\cite{Pines1957} who derived an expression
\begin{equation}\label{eqn:overhauser}
T_X = \frac{1}{\Gamma_X}=
\frac{4\pi\hbar^2s^5\rho}{\omega_0^2kT_\mathrm{res}\gamma^2IA^2}
\end{equation}
where $s$ is the sound velocity of silicon, $\rho$ is the mass
density of silicon, $\gamma$ a multiplicative factor in the range
10 to 100, $I$ the nuclear spin and $A$ the hyperfine constant
while $\omega_0=g\mu_BB$ is the Larmor frequency of the electrons
with $g$ and $\mu_B$ representing the electron Land\'e--factor and
Bohr's magneton, respectively and $B$ the applied magnetic field.
It is important to note that the Overhauser relaxation process
serves to return the two spin populations $n_2$ and $n_3$ to
thermal equilibrium with the phonon reservoir, with a temperature
$T_\mathrm{res}$, which is not necessarily the same as the spin
temperature $T_\mathrm{spin}$. Due to the constant generation of
new excess charge carriers by the illumination, a steady state
will be established in which a constant density of hot electrons
persists. As these hot electrons cascade towards the lattice
temperature, they will emit phonons at a constant rate and thus
$T_\mathrm{res}>T_\mathrm{spin}$. Note that the phonons will also
increase $T_\mathrm{spin}$, however, this effect is minimal due to
the thermal mass of the silicon, which is held constant by the
helium bath. Differences between $T_\mathrm{res}$ and
$T_\mathrm{spin}$ have previously been demonstrated using
electrical injection of hot carriers\cite{Feher1959}.
Additionally, the photoexcited carriers may scatter with the bound
donor electrons \cite{Honig1978,Ghosh1992}, causing spin
relaxation. We capture this process in our rate picture by
introducing $\Gamma_A$($\Gamma_B$), the rate coefficient for
scattering between spin up (down) free electrons and spin down
(up) bound electrons.


Feher has previously discussed the effect of the phonon reservoir
temperature on the polarization of phosphorus in silicon
\cite{Feher1959}. If the two characteristic temperatures of our
system are equal, $T_\mathrm{res} = T_\mathrm{spin}$, then the
thermally (hardly) polarized equilibrium population distribution is
obtained. However, forcing $T_\mathrm{res}>T_\mathrm{spin}$ by
photoexcitation of charge carriers, we change the steady state
population distribution. The Overhauser process will try to achieve
thermal equilibrium between states $n_2$ and $n_3$ at a temperature
$T_\mathrm{res}$, and the longitudinal relaxation process will force
states ($n_1$ and $n_2$) and ($n_3$ and $n_4$) to thermal
equilibrium at temperature $T_\mathrm{spin}$. See Fig. 1(b) for a
sketch outlining this process. The result of this situation is that
the population of $n_1$ becomes much larger than the population of
all other states, resulting in a net nuclear antipolarization, since
$P = \frac{(n_1 + n_2) - (n_3+n_4)}{(n_1 + n_2) + (n_3 +n_4)}$.
Numerical modelling of this process with realistic values for
$T_\mathrm{spin}$ and $T_\mathrm{res}$ and $T_1$ indicate that
polarization near 100\% is achievable.


\begin{figure}
\centering\includegraphics[width=85mm]{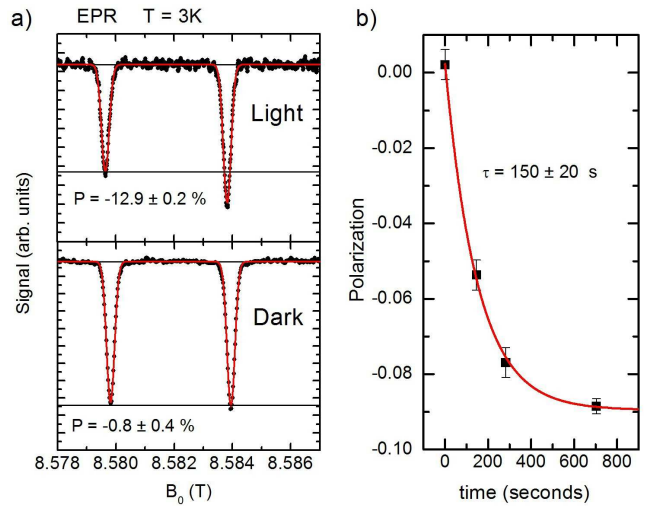}
\caption{\label{fig:EPR_Light_NoLight} (a) ESR spectra, measured at
T=3K with $f_\mathrm{res} = 240$ GHz, with (top) and without
(bottom) illumination by a mercury discharge lamp. The polarization
is determined by comparing the areas of the two resonances, obtained
by fitting the data with two Gaussian lineshapes separated by the
phosphorus hyperfine splitting, $\Delta B = 4.17$ mT (solid line).
The small magnetic field offset between the two spectra is due to
the random offset caused by the superconducting magnet. (b) The
$^{31}$P nuclear polarization measured as a function of time as
obtained from EPR spectra, at $T=3$K. The solid line represents a
single exponential decay function which was fit to the data.}
\end{figure}

To investigate this effect, we have undertaken electron spin
resonance (ESR) and electrically detected magnetic resonance (EDMR)
experiments at $B\approx 8.5$ T, corresponding to a resonant
frequency, $f = 240$ GHz \cite{Tol2005,Morley2008}. Similar
experiments have been described by us elsewhere
\cite{McCamey2008,MorleyHighField}. The samples used in this study
were similar to those described in reference \cite{McCamey2008}.
They consist of crystalline silicon with (111) surface orientation
with a phosphorus doping density $[P]\sim 10^{15}$ cm$^{-3}$, with
aluminum surface contacts to allow EDMR.

Figure \ref{fig:EPR_Light_NoLight}(a) shows two ESR spectra recorded
at $B\approx8.5$ T and $T = 3$ K. The spectra were recorded by
sweeping $B_0$ through the expected resonance fields. We fit the two
observed resonances with two gaussian line shapes. We can be sure
the signal is from phosphorus donor electrons due to both the
$g$--factor and hyperfine splitting of 4.17mT.  The low-field
(high-field) resonance is due to nuclear spins aligned
(anti-aligned) with the external field, which we will call spin up,
$\uparrow$ (spin down, $\downarrow$). The resonances are saturated
due to the long relaxation times, however, we assume that the
relaxation times are the same and, as a result, can take the area of
the resonance as a measure of the number of spins that contribute to
it \cite{Morley2007}. We thus determine the polarization of the
sample, $P = \frac{(\uparrow-\downarrow)}{(\uparrow+\downarrow)}$.
The lower spectrum was recorded in the dark, and shows a nuclear
polarization $P=-0.008\pm0.004$. Next, light from a mercury
discharge lamp was shone onto the top side of the sample through an
optical fibre, and the ESR spectra was remeasured (upper spectrum).
Again, two resonances are visible, however, they have different
intensities. Here, we determine the nuclear spin polarization $P =
-0.129\pm0.002$. This is a change in polarization over the expected
thermal polarization by a factor $\eta = P/P_0 \approx -78$. A
similar result is obtained sweeping $B_0$ in the opposite direction,
indicating that the polarization is not a passage
effect\cite{Pines1957}.

The polarization model discussed above predicts that the time taken
to reach a steady--state polarization should be limited by the
Overhauser rate, since $1/T_X = \Gamma_X << \Gamma_1, \Gamma_A,
\Gamma_B$. By using previously measured \cite{Feher1959b} low
magnetic field ($B \approx 340$ mT) values for $T_X$, and
extrapolating to the field used in the experiments presented here
using Equ.~\ref{eqn:overhauser}, we obtain for the Overhauser time
$T_X \approx 65$ s, for $T_\mathrm{res} = 3K$ and $\omega_0 = 240$
GHz. Figure \ref{fig:EPR_Light_NoLight}(b) shows the polarization
measured via ESR after light was applied to the sample. The data
shows a gradual approach to a non--equilibrium steady state. The fit
of these data with a single exponential decay function shows
excellent agreement and yields a time constant of $\tau = 150\pm20$
s. We believe this is in very good agreement with the predictions of
the Overhauser rate made by Pines et al.\cite{Pines1957}, given the
uncertainty of the low field value ($\sim 30$ hours) at a higher
donor density, and the extrapolation over nearly two orders of
magnitude of the magnetic field on which the Overhauser rate depends
quadratically.

One aspect of the experiment above suggests that the polarization
measured with ESR poses a lower limit on the maximum polarization
obtained. ESR measures the polarization in the entire sample;
however, only the surface is illuminated. We expect that, whilst
the charge carriers will diffuse throughout the sample, they will
thermalize while they diffuse. This will lead to a strong depth
inhomogeneity of the reservoir temperature and hence a depth
dependence of the polarization. While the polarization will be
biggest near the surface which is being illuminated it will be
minimized on the opposite sample surface.

\begin{figure}
\centering\includegraphics[width=8cm]{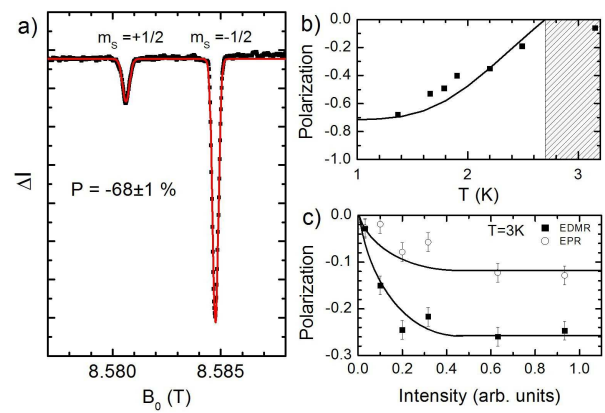}
\caption{\label{fig:EDMR} a) Electrically detected magnetic
resonance spectrum of Si:P at $T=1.37$ K under illumination. The
spectrum was fit with two gaussian lineshapes, and the nuclear
polarization determined by comparing the area of the hyperfine split
resonances. Here, $P = -68\pm1$\%. b) The polarization determined
from EDMR measurements for a number of different temperatures. The
line is a numerical simulation of the expected polarization using
the rate model presented in this paper. The shaded region indicates
where our model does not hold. See text for details. The
measurements in a) and b) were obtained using a xenon discharge
lamp. c) The polarization measured simultaneously with ESR and EDMR,
as a function of the illumination intensity of the mercury discharge
lamp used to generate the photocarriers, measured at $T=3$ K. The
lines are guides to the eye, and are scaled by a factor of
approximately 2.5 between the ESR and the EDMR.}
\end{figure}

EDMR is a magnetic resonance detection scheme which is sensitive to
spins close to the illuminated sample surface. EDMR relies on the
current through a sample being influenced by the observed spin
state. In Si:P at high magnetic fields, we have
shown\cite{MorleyHighField} that EDMR is able to be observed due to
to the spin dependent capture/emission mechanism described by
Thornton and Honig \cite{Thornton1973}, which we have included in
our polarization model with $\Gamma_A$ and $\Gamma_B$. The effect of
this process is to decrease the current through the sample when
resonant excitation of the donor electrons occurs. To measure EDMR,
we thus require free charge carriers, which are provided by the
illumination used to polarize the nuclear spins. Figure
\ref{fig:EDMR}a) shows an EDMR spectrum recorded at $T=1.37$ K, the
lowest temperature we were able to access. The spectrum was measured
with illumination by a xenon discharge lamp, and a device current,
$I_\mathrm{SD} = 500$ nA. The microwaves were chopped at a frequency
of 908 Hz, and the change in current was recorded with a lock-in
amplifier. As with the ESR measurements, the spectrum is well fit by
two gaussian lineshapes separated by the hyperfine splitting. Again,
we use the area of the resonances as a measure of the population in
each nuclear spin state. The polarization measured here is
$P=-68\pm1\%$. This corresponds to an enhancement over the
equilibrium polarization of $\eta \approx 190$, and to an effective
nuclear spin temperature of $\approx 5$ mK.

EDMR measurements allow the observation of a $^{31}$P subensemble
with a significantly more homogeneous reservoir temperature than the
ESR measurements. We therefore use EDMR to test some of the
qualitative properties of the polarization model described, namely,
the lattice temperature dependence and the illumination intensity
(and hence reservoir temperature) dependence of the observed nuclear
polarization. Fig.~\ref{fig:EDMR}b) shows the $^{31}$P polarization
as a function of the lattice temperature. It is found to increase
monotonically below $T\approx 3$ K. Based on the rate model
presented in Fig.~\ref{fig:RateDiagram}, we calculated the
polarization using the measured lattice temperature and a constant
reservoir temperature whose value was chosen to fit the experimental
data. The simulation results are also shown in Fig.
\ref{fig:EDMR}b). The best fit of the simulated values to the
measured values was achieved for $T_\mathrm{res}=2.7\mathrm{ K}$, in
agreement with the expectation that hyperpolarization vanishes when
$T_\mathrm{spin}\approx T_\mathrm{res}$. Note that there is
significant discrepancy between the fit and the data for
temperatures above $T_\mathrm{spin}=2.5$ K. While the calculated
data predicts no polarization, the measured data shows a clear
hyperpolarization of P=-6\% at $3$K. This discrepancy can be
attributed to our assumption of a constant $T_\mathrm{res}$ used in
the calculation. Note that $T_\mathrm{res}\geq T_\mathrm{spin}$ for
these experiments. Hence, the assumption of a constant
$T_\mathrm{res}\equiv 2.7$ K becomes unrealistic at
$T_\mathrm{spin}>2.7$ K.

In order to further test the polarization model we changed the
excitation spectrum of the excess charge carriers from the xenon
lamp used for the acquisition of the data in Fig.~\ref{fig:EDMR}(a)
and (b) to a mercury lamp which has a higher spectral temperature.
For the latter we measured polarization with both EDMR and ESR at a
constant lattice temperature of $T_\mathrm{spin}=3$ K. As shown in
Figure \ref{fig:EDMR}c), the EDMR spectra recorded with the mercury
lamp yield a significantly higher polarization of up to P=-24\%
(instead of 6\% at $T_\mathrm{spin}=3$ K), independently of the
intensity over a range of almost one order of magnitude. As
expected, at low intensities, when the excess charge carrier
densities drop into a range where the Overhauser process is
dominated by $T_\mathrm{spin}$, the nuclear polarization vanishes
and equilibrium appears. The polarization measured with ESR was
consistently $\approx45$ \% of that measured with EDMR, confirming
again the inhomogeneity of the reservoir temperature throughout the
sample.


Note that while we have demonstrated polarization above $P=-68$\%,
our model predicts the possibility of even higher anti--polarization
at lower temperatures and higher optical excitation rates. The
technical simplicity of this polarization method suggests that it
may be beneficial for a variety of technical applications. For
instance, silicon microparticles are biologically inert which makes
them prime candidates as contrast agents for in vivo magnetic
resonance imaging. We see no obvious reason why the polarization
technique presented above will not provide the same level of
polarization in microparticles as we have demonstrated in bulk
material. Given room temperature spin lifetimes $> 20$ minutes for
$^{31}$P nuclei in a-Si:H, a disordered material with a bigger
defect density and a larger hyperfine interaction than crystalline
silicon, we expect polarization lifetimes of over an hour for this
material, easily allowing implementation of such
experiments\cite{McCarthy1987}. Also, the rapid polarization of
$^{31}$P nuclear spins demonstrated may offer an initialization
mechanism for $^{31}$P in silicon spin qubits.


In conclusion, the data presented above demonstrates that hyper
(anti-) polarization of phosphorous donor nuclear spins in
crystalline silicon can be achieved rapidly (of the order of a few
minutes) by irradiation with above silicon bandgap light at low
temperatures and high magnetic fields. Polarization in excess of
68\% was demonstrated, and discussed in terms of a model arising
from the increased reservoir temperature driven by phonon emission
during thermalization of photoexcited carriers. The qualitative
predictions of this model for the polarization dependence on lattice
temperature, illumination temperature and intensity have been
verified and technical applications of this effect have been
discussed.

This work was supported by a Visiting Scientist Program Grant
7300-100 from the National High Magnetic Field Laboratory. GWM was
supported by the EPSRC through grants GR/S23506 and EP/D049717/1.

\end{document}